# Quark beta decay in an inhomogeneous chiral phase and cooling of hybrid stars

**T. Tatsumi**
*Department of Physics, Kyoto University*
*Kyoto 606-8502, Japan*
*E-mail:* `tatsumi@ruby.scphys.kyoto-u.ac.jp`

**T. Muto**
*Department of Physics, Chiba Institute of Technology*
*2-1-1 Shibazono, Narashino, Chiba 275-0023, Japan*
*E-mail:* `takumi.muto@it-chiba.ac.jp`

We discuss the cooling of hybrid stars by considering the neutrino emission from quark matter. As a current topic the appearance of various inhomogeneous chiral phases have been studied near the chiral transition. Here we consider the dual-chiral-density-wave (DCDW) specified by the spatially modulated quark condensates. Since the DCDW state can be represented as a chirally rotated state from the normal quark matter, the quark weak-current is accordingly transformed to have an additional phase factor which modifies the energy-momentum conservation at the vertex, and makes the quark direct Urca process possible. The direct evaluation of the neutrino emissivity shows that it is proportional to $T^6$ and the magnitude is comparable with the quark or pion cooling. Since the DCDW phase develops only in the limited density region, this novel mechanism may give an interesting scenario about cooling of hybrid stars that lower-mass stars should be cooler than higher-mass ones.

# 1. Introduction

Chiral transition in QCD phase diagram has been discussed by many authors. Nowadays the appearance of the inhomegeneous phases is an important topic near the chiral transition. Raising the chiral order parameter s.t.

$$\Sigma \equiv \langle \bar{q}q \rangle + i \langle \bar{q}i\gamma_5\tau_3 q \rangle = \Delta \exp(i\theta) \in \mathbb{C}, \qquad (1)$$

inhomogeneous phases are specified by the spatial modulation of $\Delta$ or $\theta$, while the usual chiral transition is described by the uniform $\Delta$ with $\theta = 0$. Two phases have been known as typical examples: dual-chiral-density-wave (DCDW) with $\Delta = $ constant and $\theta = \mathbf{q}\cdot\mathbf{r}$ [1], and real kink crystal (RKC) with $\Delta(\mathbf{r})$ and $\theta = 0$ [2]. In the model calculations using NJL model the chiral critical point (CCP) has been suggested on the $\mu - T$ plane. It has been shown by the Ginzburg-Landau approach that once the inhomogeneous phases are taken into account, CCP should give away to the Lifshitz point [2]. Here we consider an implication of such inhomogeneous phases on compact star phenomena, especially cooling, by taking DCDW.

Cooling of compact stars have provided information about form of matter at high-densities [3,4]. Recent observation of the surface temperature of young pulsars have suggested interesting possibilities: (1)3C58 or Vela looks to have rather low temperature which should be hard to be explaiened by the standard scenario. (2) Cas A is hot in spite of its large mass [5]. The first point has been a long-standing problem, and someone suggested *exotic cooling mechanisms* to explain them (fast cooling scenario) [3]. As an exotic cooling mechanism we can consider the direct Urca process in quark matter (quark cooling), which is a most efficient cooling mechanism in hybrid stars,

$$d \rightarrow u + e^- + \bar{\nu}_e, \quad u + e^- \rightarrow d + \nu_e \qquad (2)$$

For free massless quarks, this process is strongly prohibited at low temperature due to the kinematical condition (*triangular condition*). However, once the one-gluon-exchange (OGE) interaction is taken into account a la Landau Fermi-liquid theory, momentum triangle can be formed among quarks and electron to give the emissivity $\varepsilon \propto \alpha_s T^6$ with the QCD coupling constant $\alpha_s$ [6]. So it has a perturbative nature as it is.

Although this possibility is interesting in the light of recent theoretical development, fast cooling is always switched on once density exceeds the critical density. So it looks to contradict the observation of Cas A. Thus we need further idea to explain the cooler stars and Cas A in a consistent way. We shall show the quark Urca process in the presence of DCDW satisfies these requirements to give a large neutrino emissivity like other exotic cooling mechanisms in the limited density region.

# 2. Emergence of DCDW in quark matter

Consider the NJL model with two flavors,

$$L_{\text{NJL}} = \bar{\psi}\left(i\partial\!\!\!/ - m_c\right)\psi + G\left[\left(\bar{\psi}\psi\right)^2 + \left(\bar{\psi}i\gamma_5\boldsymbol{\tau}\psi\right)^2\right] \tag{3}$$

in the chiral limit $m_c = 0$. Leaving the neutral mean-fields and taking the form for DCDW, $\Sigma = \Delta \exp(i\mathbf{q}\cdot\mathbf{r})$, we can see the scalar and pseudoscalar condensates spatially modulate along one direction [1]. Note that phase degree of freedom is important for DCDW. Introducing the Weinberg transformation, $\psi_W = \exp[i\gamma_5\tau_3\mathbf{q}\cdot\mathbf{r}/2]\psi$, we find that the single-particle energy can be written as

$$E^{\pm}(\mathbf{p}) = \sqrt{E_p^2 + |\mathbf{q}|^2/4 \pm \sqrt{(\mathbf{p}\cdot\mathbf{q})^2 + M^2|\mathbf{q}|^2}}, \quad E_p = \left(|\mathbf{p}|^2 + M^2\right) \tag{4}$$

with the dynamical mass $M = -2G\Delta$, which exhibits the splitting of the energy, depending on the "spin" degree of freedom like the exchange splitting in the condensed matter physics. Accordingly the Fermi surfaces are deformed as $|\mathbf{q}|$ changes; the Fermi surface for the majority has a prolate shape while the one for the minority a oblate shape for $\mathbf{q} \parallel \hat{z}$. Some numerical results are presented in Fig. 1 [1], which shows the appearance of DCDW on the region between $\mu_{c1}$ and $\mu_{c2}$.

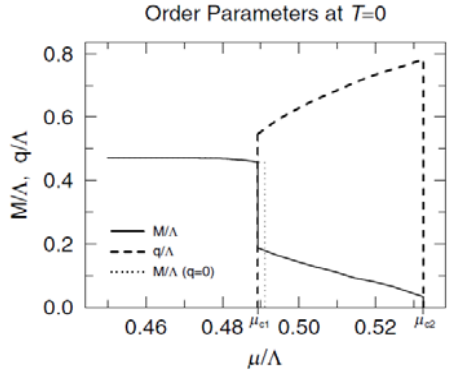

Figure1

### 3. Quark beta decay in the DCDW phase

We consider the quark beta decay in the DCDW phase, $d \to u + e^- + \bar{\nu}_e$. First, note that the ground state can be represented as a chirally rotated state,

$$|\text{DCDW}\rangle = \exp\left(i\mathbf{q}\cdot\int \mathbf{r}A_3^0(\mathbf{r})d^3r\right)|\text{normal}\rangle \left(\equiv U_{\text{DCDW}}(\mathbf{q}\cdot\mathbf{r})|\text{normal}\rangle\right) \tag{5}$$

The effective Hamiltonian is given as the current-current form, $H_W = \frac{G_F}{\sqrt{2}} h^\mu_{1+i2} l_\mu + h.c.$ with $h^\mu_{1+i2} = V_1^\mu + iV_2^\mu - A_1^\mu - iA_2^\mu$, so that the transition matrix element is given as $\langle u, e^-, \bar{\nu}_e | H_W | d \rangle = \langle u_W, e^-, \bar{\nu}_e | \tilde{H}_W | d_W \rangle$, where

$$\tilde{H}_W = U_{\text{DCDW}} H_W U_{\text{DCDW}}^\dagger = \frac{G_F}{\sqrt{2}} \tilde{h}^\mu_{1+i2} l_\mu + h.c.. \tag{6}$$

The transformed quark current $\tilde{h}^{\mu}_{1+i2}$ now reads

$$\tilde{h}^{\mu}_{1+i2} = U_{\text{DCDW}} h^{\mu}_{1+i2} U^{\dagger}_{\text{DCDW}} = \exp(i\mathbf{q}\cdot\mathbf{r}) h^{\mu}_{1+i2} \tag{7}$$

by way of the current algebra, which clearly implies that DCDW modifies the momentum conservation at the weak-interaction vertex. So the triangular condition is easily satisfied without OGE interaction. Here it should be interesting to compare it with the pion cooling [7,8].

Consider the following beta decay process: $d(p_1) \rightarrow u(p_2) + e^-(p_3) + \bar{\nu}_e(p_4)$. The neutrino emissivity then can be given as

$$\varepsilon = 2N_C G_F^2 \int \frac{d^3 p_3}{(2\pi)^3 2E_3} \int \frac{d^3 p_4}{(2\pi)^3 2E_4} E_4 L_{\lambda\alpha} n_F(-E_3 + \mu_e) f_B(E_4 - \mu_e + E_3) \text{Im} \Pi_R^{\lambda\sigma}(k), \tag{8}$$

with $k = (E_3 + E_4 - \mu_e, \mathbf{p}_3 + \mathbf{p}_4)$, where $n_F$ and $f_B$ are the Fermi-Dirac and Bose-Einstein distribution functions, respectively, and the factor 2 accounts for the neutrino process [9]. The information of the quark tensor is summarized in the W boson polarization tensor,

$$\Pi_R^{\lambda\sigma}(k) = T \sum_n \int \frac{d^3 p_1}{(2\pi)^3} \text{tr}\left[\Gamma^{\lambda} S_W^d(p_1) \Gamma^{\sigma} S_W^u(p_1 - k \pm q)\right] \tag{9}$$

with $\Gamma^{\mu} = \gamma^{\mu}(1-\gamma_5)$ and quark propagator, $S_W^{-1}(p) = \slashed{p} - M + \gamma_5 \tau_3 \slashed{q}/2$. $L_{\lambda\alpha}$ is the standard leptonic tensor. After some manipulation we have a final expression for the emissivity

$$\varepsilon = 6V^{-1}\left[\prod_{i=1}^{4} V \int \frac{d^3 p_i}{(2\pi)^3}\right] E_4 W_{fi} n_d(\mathbf{p}_1)(1 - n_u(\mathbf{p}_2))(1 - n_e(\mathbf{p}_3)), \tag{10}$$

where the transition rate $W_{fi}$ is given by the product of the leptonic tensor and quark tensor $H_{\alpha\beta}$, $H_{\alpha\beta} = \text{tr}(\rho_u \Gamma_{\alpha} \rho_d \Gamma_{\beta})$ in terms of the density matrix $\rho_f$. Note that the quark tensor is now complicated due to the propagator in the presence of DCDW. The phase-space integral is also complicated due to the anisotropy of the momentum distribution. Thus we must numerically calculate the emissivity (10) in the general case. Before discussing the definite value of the emissivity, it should be worth mentioning that the counting of $T$ dependence goes in the same way as the quark or pion cooling to give rise to $\varepsilon \propto T^6$.

## 4. Results

We roughly estimate the emissivity for two regions: one is near the onset density $\mu_{c1}$ and the other is near the termination density $\mu_{c2}$.

### 4.1. Near the onset density ($\mu \simeq \mu_{c1}$)

There appear two Fermi surfaces, depending on the polarization, for each flavour; the volume of the majority is already much bigger than that of the minority, so that we may safely discard the minority for simplicity. On the other hand, the deformation of the Fermi surfaces are still small at the onset density, one may introduce "equivalent Fermi spheres" to roughly estimate the emissivity (Fig. 2); preserving the whole volume (the number), we find the effective Fermi momentum $\bar{p}_{Fi} \simeq \mu_i + q/2$. Then momentum conservation can be expressed by $\bar{\mathbf{p}}_d = \bar{\mathbf{p}}_u + \mathbf{p}_e \pm \mathbf{q}$. Single particle energy is also approximated as $E_p^\pm \simeq |\mathbf{p}|$ since $\mu_i >> M \simeq q/2$ (see Fig. 1).

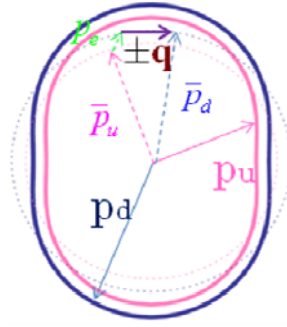

Figure 2

Then the squared matrix element $W_{fi}$ is reduced to a simple form, and we finally find

$$\varepsilon_{\text{DCDW}} = \frac{457}{1680} \pi G_F^2 \mu_d \mu_u \frac{\mu_e^2}{q} T^6 \qquad (11)$$
$$\simeq 6 \times 10^{26} (\rho/\rho_0) Y_e^{2/3} T_9^6 (\text{ergcm}^{-3}\text{s}^{-1}),$$

where $T_9$ is the temperature in units of $10^9$ K. Note that we implicitly assumed that triangular condition is satisfied. In the limit $q \to 0$ it is not obviously satisfied..

**4.2 Near the termination density** $(\mu \simeq \mu_{c2})$

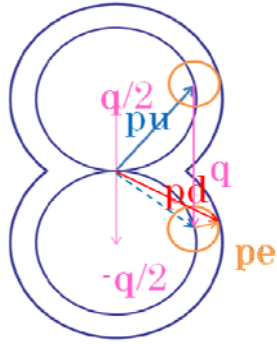

Figure 3

The Fermi surfaces are well deformed in this case, which suggests that they can be approximated by two separated spheres with centers shifted by $q$ (Fig. 3). Single-particle energy now renders $E_p^\pm \simeq |\mathbf{p} \pm \mathbf{q}/2|$, since the dynamical mass $M$ is almost vanished there. The momentum conservation simply reads $\mathbf{p}_d = \mathbf{p}_u + \mathbf{p}_e \pm \mathbf{q}$. The final result is

$$\varepsilon_{\text{DCDW}} = \frac{3\pi}{4} \frac{457}{5040} G_F^2 \frac{\mu_d}{\mu_u} M^2 |\mathbf{q}| \ln\left[\frac{2|\mathbf{q}|}{M}\right] T^6 \qquad (12)$$
$$\simeq 10^{25} T_9^6 (\text{ergcm}^{-3}\text{s}^{-1}).$$

There appears a logarithmic dependence on *q* or *M* due to a singular structure of the quark tensor.

**5. Concluding remarks**

We have proposed a novel cooling mechanism (DCDW cooling) of hybrid stars, based on the idea of the inhomogeneous chiral phase in quark matter. It comes from the non-perturbative effect of QCD at moderate densities. We have seen that DCDW modifies the momentum conservation at the weal-interaction vertex by supplying an additional momentum. The emissivity has a $T^6$ dependence and should be very efficient for young pulsars. We evaluated the emissivity in two density regions to find that the magnitude is $O(10^{25-26} T_9^6 (\mathrm{ergcm}^{-3}\mathrm{s}^{-1}))$ and comparable with those for other exotic cooling mechanisms such as pion or quark cooling. Numerical calculation is needed in the general case.

Since the inhomogeneous phase should be limited to the vicinity of the chiral transition, DCDW cooling works in the narrow density region. Thus the cooling source may be distributed in the shell region inside hybrid stars. Assuming that usual quark cooling is strongly suppressed by color superconductivity and DCDW cooling works instead, one may say that heavier stars do not necessarily cool fast [10].

Realistic EOS of the DCDW phase is needed for further study, where charge neutrality, chemical equilibrium or symmetry breaking effect should be taking into account. Effects of the magnetic field is an interesting subject, since the appearance of the DCDW should be a robust conclusion in the presence of the strong magnetic field [11], independent of details of the definite models.